# A stable and low loss oxide layer on α-Ta (110) film for superconducting qubits

Zengqian Ding[1], Boyi zhou[1], Tao Wang[1],Lina Yang[1], Yanfu Wu[1], Xiao Cai[1], Kanglin Xiong[1,2,a)], Jiagui Feng[1,2,b)]

[1] Gusu Laboratory of Materials, Suzhou, 215123

[2] Vacuum Interconnected Nanotech Workstation (Nano-X), Suzhou Institute of Nano-Tech and Nano-Bionics, CAS, Suzhou, 215123

[a)] E-mail: klxiong2008@sinano.ac.cn

[b)] E-mail: jgfeng2017@sinano.ac.cn

## ABSTRACT

The presence of amorphous oxide layers can significantly affect the coherent time of superconducting qubits due to their high dielectric loss. Typically, the surface oxides of superconductor films exhibit lossy and unstable behavior when exposed to air. To increase the coherence time, it is essential for qubits to have stable and low dielectric loss oxides, either as barrier or passivation layers. In this study, we highlight the robust and stable nature of an amorphous tantalum oxide film formed on α-Ta (110) by employing chemical and structural analyses. Such kind of oxide layer forms in a self-limiting process on the surface of α-Ta (110) film in piranha solution, yielding stable thickness and steady chemical composition. Quarter-wavelength coplanar waveguide resonators are made to study the loss of this oxide. One resonator have a $Q_i$ of $3.0\times10^6$ in the single photon region. The $Q_i$ of most devices are higher than $2.0\times10^6$. Moreover, most of them are still over one million even after exposed to air for months. Based on these findings, we propose an all-tantalum superconducting qubit utilizing such oxide as passivation layers, which possess low dielectric loss and improved stability.

# I. INTRODUCTION

Superconducting quantum computing has gained substantial experimental progresses over the last two decades. It is now one of the leading candidates to build a fault-tolerant quantum computer.[1] F. Arute et al. achieved quantum "supremacy" in 2019 with a chip of 53 X-mon qubits.[2] To demonstrate real quantum advantage, the quality and quantity of qubits need to increase simultaneously. In recent years, the coherence time ($T_1$) of Transmon qubits on plane surface has increased to hundreds of microseconds due to the improvement of material quality and fabrication methods.[3-7] In 2020, A. P. M. Place et al. reported coherence time of exceeding 0.3 ms for the Transmon fabricated with alpha-phase tantalum films.[8] Following this report, C. Wang et al. increased the coherence time to 0.5 ms in 2021.[9]

It is believed that most lossy channels emerged in the material growth and device fabrication processes.[10,11] Specifically, the dielectric loss arises from three interfaces,[12] i.e., the amorphous oxide layer at the superconductor-air (MA) interface, the amorphous oxide layer at the substrate-air (SA) interface, and the amorphous layer at the superconductor-substrate (MS) interface. Substrate cleaning, epitaxial film growth and subsequently etching method for the qubit fabrication can improve the MS interface.[13-17] However, the amorphous oxide layers in the MA interface is inherent with material properties and hard to manage. Most superconductors used in the qubits are metals which are easy to be oxidized in air, and the thicknesses of these oxides increase over time. Sustained efforts have been paid to reduce loss of the MA interface. Verjauw et al. discovered that removing the niobium (Nb) oxides can increase the internal quality factor ($Q_i$) with half magnitude.[18] Other groups also found that passivation of the superconductor surface can improve the quality of the devices.[19] Furthermore, because of the excellent performance of α-Ta film devices, the research and surface treatment of Ta-films have progressed.[20-22] Even though the detailed analysis of the chemical composition of the amorphous oxide layer mainly focuses on the α-Ta (111) films on C-plane sapphire substrates.[22] These studies show that the surface oxides of Ta film are a major source of two level system (TLS) losses, especially some of these sub-oxides, and that these losses can be reduced by chemical etching.[21,23] Besides the three interfaces, the Josephson junction can also be lossy and noisy. The best insulator layer in the Josephson junction is still amorphous alumina, which is unstable and has large dielectric loss.

In this paper, we investigate the surface oxide of α-Ta (110) films prepared in piranha. As to the α-Ta (110) film, the oxide layer thickness is about 2.24 nm, containing not only pentavalent tantalum ($Ta^{5+}$), but also trivalent tantalum ($Ta^{3+}$) after exposing to air. With the piranha solution treatment, the surface oxide layer is mainly composed of pentavalent tantalum ($Ta_2O_5$), with an oxide layer thickness of about 2.61 nm. Multiple piranha treatments result in little change in thickness. The CPW resonators are made of α-Ta (110) film with this kind of $Ta_2O_5$ surface oxide. The highest $Q_i$ is about three million in the single-photon regime at 10 mK. The $Q_i$ of most devices are higher than $2.0\times10^6$. Moreover, most of them are still over one million even after exposed to air for months. The high $Q_i$ and robustness of the resonators indicate that such $Ta_2O_5$ layer is of low dielectric loss and stable. Thus, we suggest an all-tantalum superconducting qubit incorporating such oxide as passivation layers, which exhibit low dielectric loss and enhanced stability.

# II. EXPERIMENTAL

The α-Ta (110) films with thickness of 200 nm were deposited on 2-inch C-plane sapphire substrates by DC magnetron sputtering in a high vacuum chamber. Prior to the film deposition, the substrate was thermally cleaned inside the sputtering chamber at 700 ℃ for 30 minutes, and then slowly cooled down to 400 ℃ at a rate of 30 ℃ per minute. During the optimal deposition process, the substrate was maintained at 400 ℃. The sputtering pressure of Ar gas was kept constant at 15 mTorr, and the DC sputtering power was set to 200 W. X-ray Diffraction (XRD) confirmed the film crystallographic plane is shown in Fig. S1 in the supplementary material [URL will be inserted by AIP Publishing]. For comparison, four different α-Ta (110) samples from the same wafer were prepared, i.e., one untreated with piranha, two soaked in piranha for 1 and 3 times and 20 minutes each time, and one exposed to air for 4 months. The piranha solution in our experiment is a mixture of sulfuric acid and 30% hydrogen peroxide in a volume ratio of 2:1. Surface morphology and roughness are measured by atomic force microscopy (AFM). The thicknesses and chemical compositions of oxide layers are investigated using scanning transmission electron microscope (TEM/STEM) and angle-resolved X-ray photoelectron spectroscopy (XPS). The XPS spectrum at different angles is presented in Fig. S2 in the supplementary material [URL will be inserted by AIP Publishing]. $Q_i$ of the coplanar waveguide (CPW) resonators made from α-Ta (110) film are measured at 10 mK. The measurement configuration is shown in Fig. S3 in the supplementary material [URL will be inserted by AIP Publishing]. The input signal path incorporates a variable attenuator at room temperature, a 20 dB attenuator at 4 K, a 6 dB attenuator at the still plate, and a 30 dB attenuator and an infrared filter at 10 mK. At 4 K on the output signal path, a HEMT amplifier with a gain of 40 dB and a room temperature amplifier with a gain of 40 dB are employed. The output signal path also incorporates three isolators and a low-pass filter. The test chips are encapsulated in high-purity Al-made sample boxes with a μ-metal magnetic shielded. The measurement is done by a vector network analyzer (VNA) with variable output power. The excitation power is applied from -160 dBm to -100 dBm by adjusting the output power of VNA and the variable attenuator. The photon number in the CPW resonators can be estimated from the input power of the pump using the following equation:

$$< n > \approx \frac{Q^2 \cdot P_{in}}{\pi^2 h f^2 \cdot Q_c}, \tag{1}$$

$$\frac{1}{\mathrm{Q}} = \frac{1}{\mathrm{Q_i}} + \frac{1}{\mathrm{Q_c}}. \tag{2}$$

where $n$ is the estimated photon number, $P_{in}$ is the pump power, $h$ is Planck's constant, $f$ is the resonance frequency of the CPW resonators, $Q$ is the loaded (total) quality factor of the resonator, $Q_c$ is the value which is related to the coupling capacitance between the CPW resonator and driveline, and $Q_i$ is the intrinsic quality factors. In order to ensure the acquisition of high-quality $Q_i$, the designing value of $Q_c$ was set at 300,000.

Due to the use of wet etching process, there will be some differences in the actual devices. The $Q_i$ was extracted by fitting the $S_{21}$ vs. frequency curve. The corresponding fitting formula can been seen in section IV the supplementary material [URL will be inserted by AIP Publishing]. Moreover, an example of a fitting curve is shown in Fig. S4 in the supplementary material [URL will be inserted by AIP Publishing].

# III. RESULTS AND DISCUSSION

The surface of as grown α-Ta (110) film with optimal deposition condition is shown in Fig. 1(a). The grains of Ta are elongated with tetragonal symmetry and tightly packed given a surface roughness (Rq) of 1.09 nm in the area of 2 μm × 2 μm. After exposed to air for hours, the film surface is covered with an amorphous oxide layer uniformly, with no sign of further oxidation of grain boundary under TEM. The thickness of this layer is 2.24 nm [Fig. 1(d)], which increases to 2.41 nm after leaving the film in air for another four months. Conversely, with immersion in piranha solution for one time, the thickness of the amorphous oxide layers increases to 2.61 nm and the surface remains alike [Fig. 1(b) and (e)]. Then exposing the sample to air and immersing the sample in piranha solution again for more time, there is no noticeable change of its surface and thickness [Fig. 1(c) and (f)]. The results suggest that the oxidization of the α-Ta (110) surface in air and piranha are self-limiting processes which probably evolve logarithmically with time.[24,25] The difference is that, in air, the surface oxide layer grows slowly and takes longer time to saturate; but in the piranha, which is a strong oxidizer, the surface oxide layer reaches its limited thickness in a shorter time.

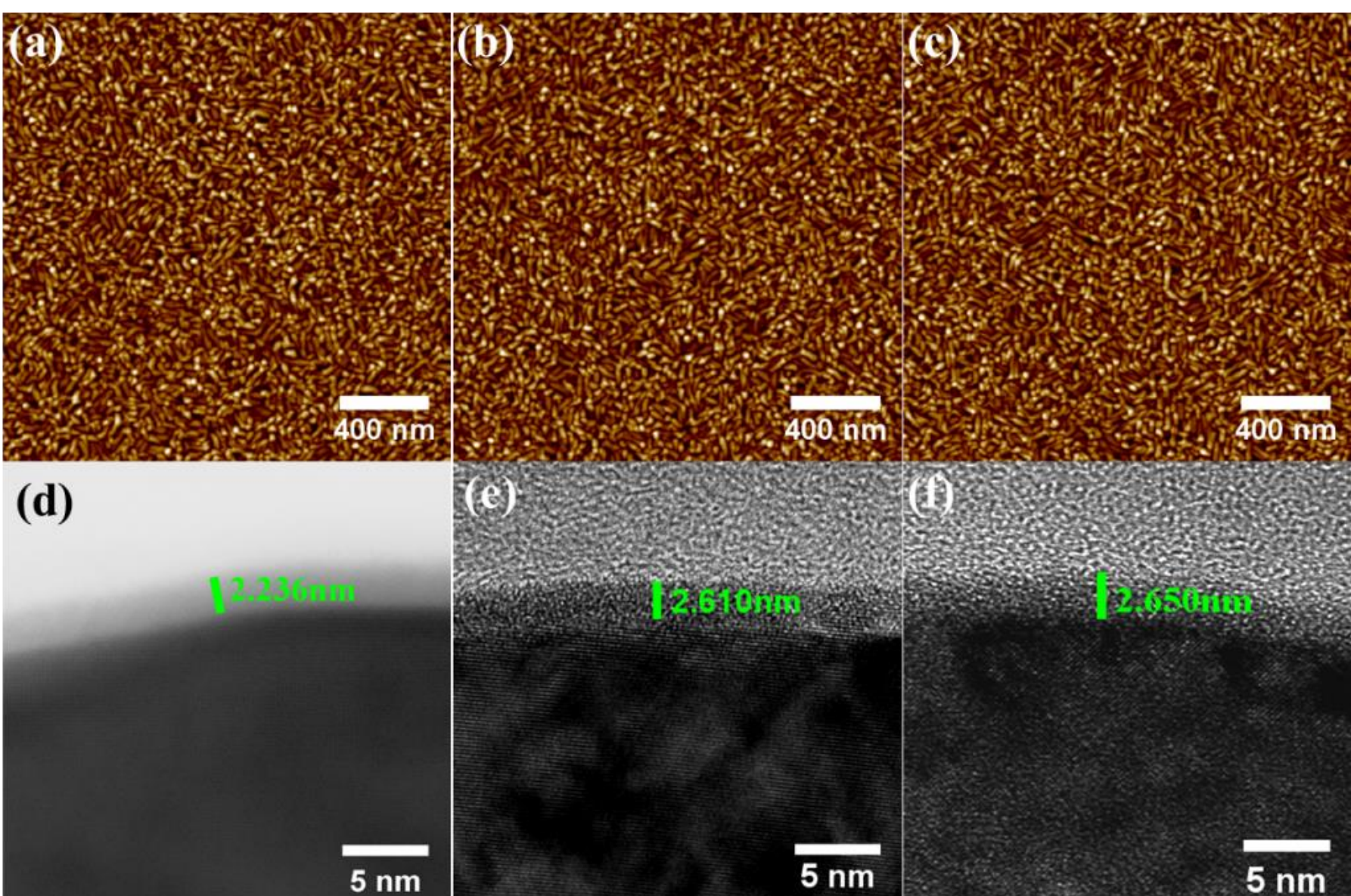

FIG. 1. AFM images of α-Ta (110) films measured over a 2 μm×2 μm area, and TEM/STEM images of Ta film surface oxide layer. (a) - (c) AFM images of α-Ta (110) film before immersion in piranha solution and immersion in piranha solution once, and three times, respectively. (d) – (f) The oxide layer thickness of the α-Ta (110) film before immersion in piranha solution, immersion in piranha solution once, and three times, respectively.

The chemical composition of the α-Ta (110) surface oxide layer is measured by the XPS. Fig. 2(a)-(b) shows the XPS data obtained at angle of 0° for two α-Ta (110) samples with oxides grown in air and piranha, respectively. Compared with the oxide grown in piranha solution (Blue curve in Fig. 2(b)), the oxide grown in air (Orange curve in Fig. 2(a)) has an obvious shoulder on the right of the two peaks at lower binding energy which are correspond to the orbitals of sub-oxidation states, as shown in the corresponding fitting curves. This is further confirmed by the angle-resolved testing as shown in Fig. 2(c). The oxide layer of α-Ta (110) film immersed in piranha solution contains only pentavalent tantalum ($Ta^{5+}$) from surface to subsurface. However, with the detecting depth increasing, the oxide layer of α-Ta (110) film exposed to air start to contain sub-oxidation states. These states can be fitted with the trivalent tantalum ($Ta^{3+}$).

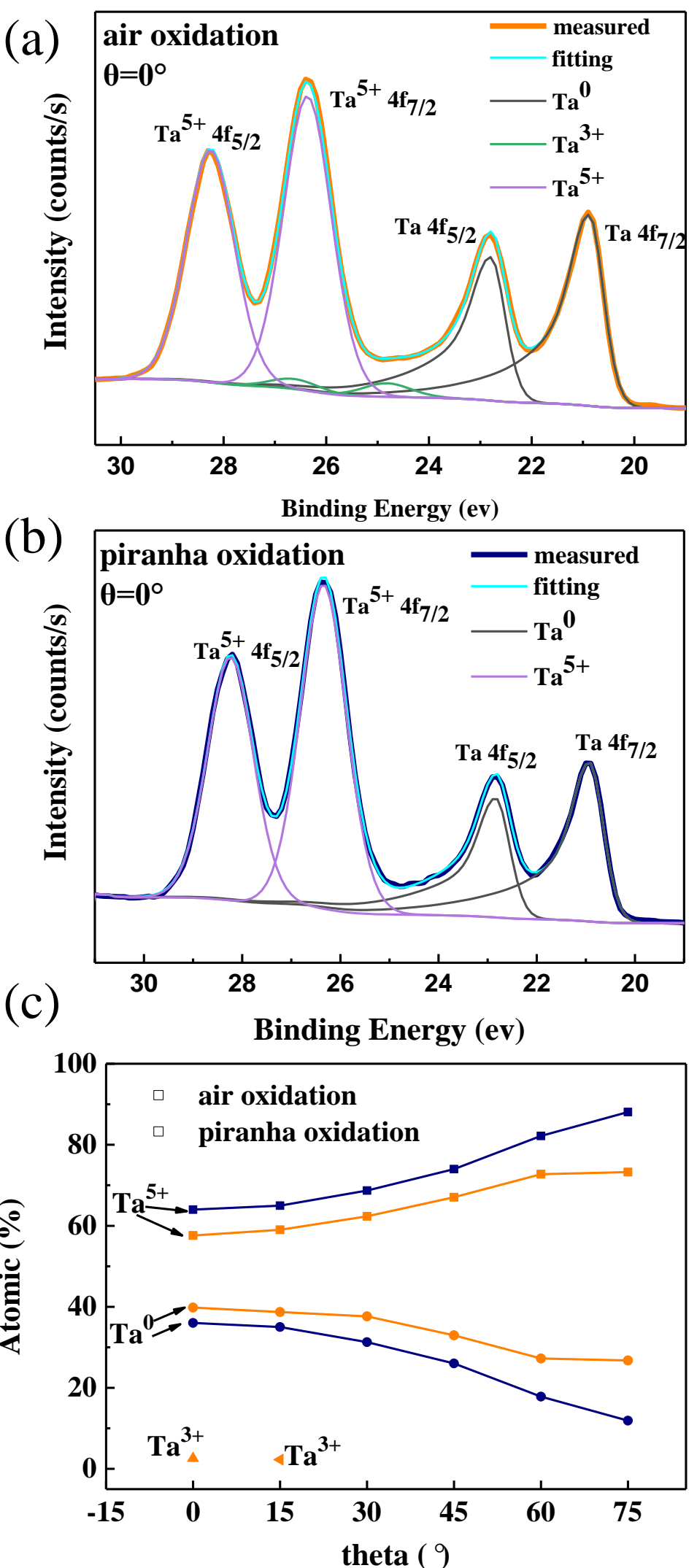

FIG. 2. Angle-resolved XPS spectra of α-Ta (110) film and the chemical elements atomic ratio in the oxide layer. The θ is the angle between the normal of the film surface and the signal acceptance angle. (a) XPS fitting data of α-Ta (110) oxidized in air at θ =0°. (b) XPS fitting data of α-Ta (110) oxidized in piranha at θ=0°. The two peaks at low binding energy belong to $4f_{7/2}$ and $4f_{5/2}$ orbitals of Ta (110) metal, and the two peaks at higher binding energy correspond to the same orbitals of $Ta_2O_5$. (c) Surface chemical elements atomic ratio according to angle-resolved XPS spectra in the oxide layer of α-Ta (110) film exposed to atmosphere (Orange curve) and immersed in piranha solution (Blue curve), respectively.

Considering the factor that the oxide prepared in piranha is almost pure $Ta_2O_5$ in a dense amorphous structure, no wonder it is very stable in air and in piranha solution. From the angle-resolved data,[26] the thicknesses of the surface oxide layer are also obtained. They are 2.45 nm for the one exposed to the air and 2.74 nm for the one immersed in piranha solution, which is consistent with the TEM data, with deviation of about 0.20 nm. The agreement of thicknesses measured by two distinct methods suggest again that the surface oxide on α-Ta (110) film formed in piranha solution is uniform.

As to the superconducting device, the metallic and semiconductor nature of sub-oxides lead to conductivity losses.[5] Since the losses as a whole can be estimated from $Q_i$ of resonators, we fabricated CPW resonators using the α-Ta (110) film covered with such oxides across different positions on the wafer scale. The devices were designed with center conductor and insulating gap widths of $w$ = 10 μm and $g$ = 6 μm, respectively. Fig. S5 and Fig. S6 in the supplementary material [URL will be inserted by AIP Publishing] show the dimensions and etching edge structure of the resonator. The resonance frequency was 6.5 GHz. To begin with, the sapphire wafer with α-Ta (110) film was soaked in piranha. The resonators were defined with optical lithography and a wet etching process. Then, the wafer was cut into 8 mm × 8 mm square chips. Last, the chips were rinsed in piranha again.

The $Q_i$ of the new resonators and resonators these had been exposed to air for four months are measured and shown in Fig. 3. One resonator has a $Q_i$ of $3.0\times10^6$ in the single photon region. This value is lower than that in the Ref. 23. The observed discrepancy can be attributed to several factors. Firstly, the CPW resonator design could be distinct. Secondly, the use of pure Al sample boxes for packaging the test chips may result in a higher actual sample temperature compared to the displayed temperature of the dilution refrigerator. Furthermore, the absence of infrared shielding covers for the test samples could also contribute to the discrepancy. From the participation ratio analysis, the up limited loss tangent of the amorphous $Ta_2O_5$ is about $1.5\times10^{-2}$. The relevant calculation formulas and values of dielectric constants are provided in Section VII of the supplementary materials [URL will be inserted by AIP Publishing]. The $Q_i$ of most devices are higher than that ($2.0\times10^6$) of the resonator made from an aluminum (Al) film deposited by molecular beam epitaxy on sapphire substrate,[14] and that ($1.0\times10^6$) of the resonator made from a Nb film deposited by sputtering on silicon substrate.[18] Regarding the surface roughness (Such as 0.40 nm for Al/Sapphire and 0.58 nm for Nb/Si) in these references and 1.09 nm for the α-Ta (110)/Sapphire here, it is fair to infer that $Ta_2O_5$ layer has a lower dielectric loss than that of the amorphous oxide layer formed in air on Al and Nb film surfaces. For the resonators left in air for four months, a slight degradation is observed. In view of many factors, like surface contamination, can substantially plague the resonators, the $Q_i$ above million is quite remarkable. The robustness of the resonators further imply that the amorphous $Ta_2O_5$ layer is stable in atmosphere.

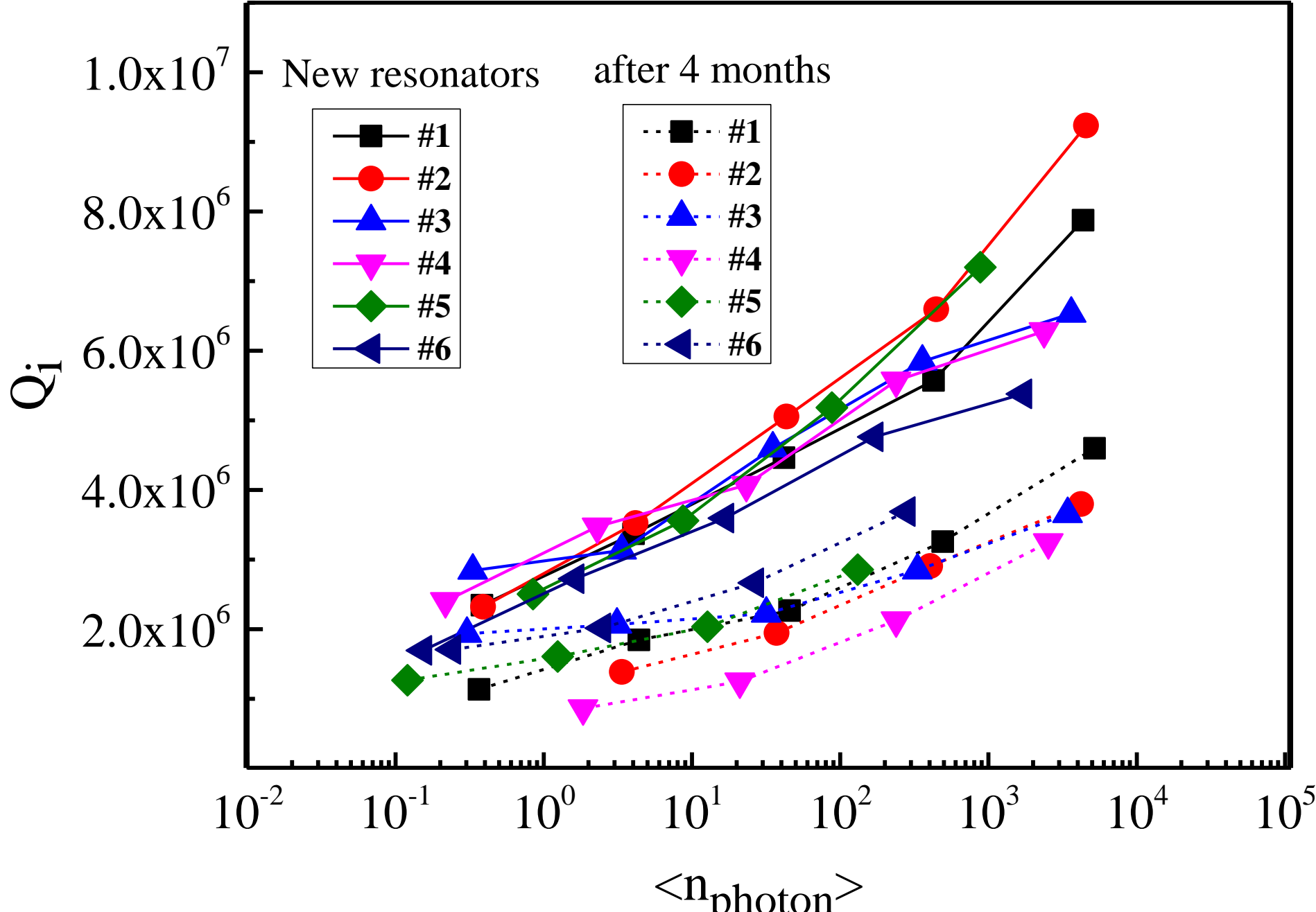

FIG. 3. The $Q_i$ vs. photon number curves of series of resonators made from α-Ta (110) film in different area on wafer scale measured immediately and after 4 months.

With the stability and low loss properties of the $Ta_2O_5$, an all-tantalum superconducting qubit using the $Ta_2O_5$ as the dielectric and passivation layer is proposed. Fig. 4 shows the technique to fabricate the Ta/$Ta_2O_5$/Ta Josephson junctions (JJs), and outlines the key fabrication steps for qubits. Before depositing Ta, substrates such as high resistivity silicon or sapphire are pretreated to ensure the clean surface and minimize dielectric loss [Fig. 4(a)]. Then, a quasi-epitaxial α-Ta (110) layer with high crystallinity and smooth surface is deposited on the substrate [Fig. 4(b)]. The Ta layer is the bottom electrode of JJs, as well as the foundation of the tunnel barrier layer. Fig. 4(c) gives the $Ta_2O_5$ tunnel barrier layer. There are several approaches that can be attempted to prepare this $Ta_2O_5$ amorphous layer, including direct piranha solution treatment combined with vacuum in-situ oxygen plasma treatment, plasma-enhanced atomic layer deposition, and radio frequency bias oxygen plasma treatment of the α-Ta (110) film under an oxygen atmosphere. Then another α-Ta (110) layer is deposited on $Ta_2O_5$ layer as the top electrode layer [Fig. 4(d)]. The pattern of JJs is defined using photolithography [Fig. 4(e)], and dry etching process is suggested because of its stable and highly anisotropic characteristic.[9] The etched surface is passivated to protect the whole JJs and other large area structures [Fig. 4(f)]. Piranha solution treatment is recommended because an amorphous $Ta_2O_5$ layer can cover all structure with new surfaces. Additionally, piranha solution can remove organic impurities, such as residual photoresist. Then, as shown in Fig. 4(g), large area structures such as resonators and capacitors are patterned into the bottom Ta layer to break the junction from surroundings. The surfaces are passivated again with piranha solution [Fig. 4(h)]. The final connection between the top electrode of the JJs and the other structure can be achieved through the use of air bridges. The process is CMOS compatible and provides scalability for superconducting quantum chips.

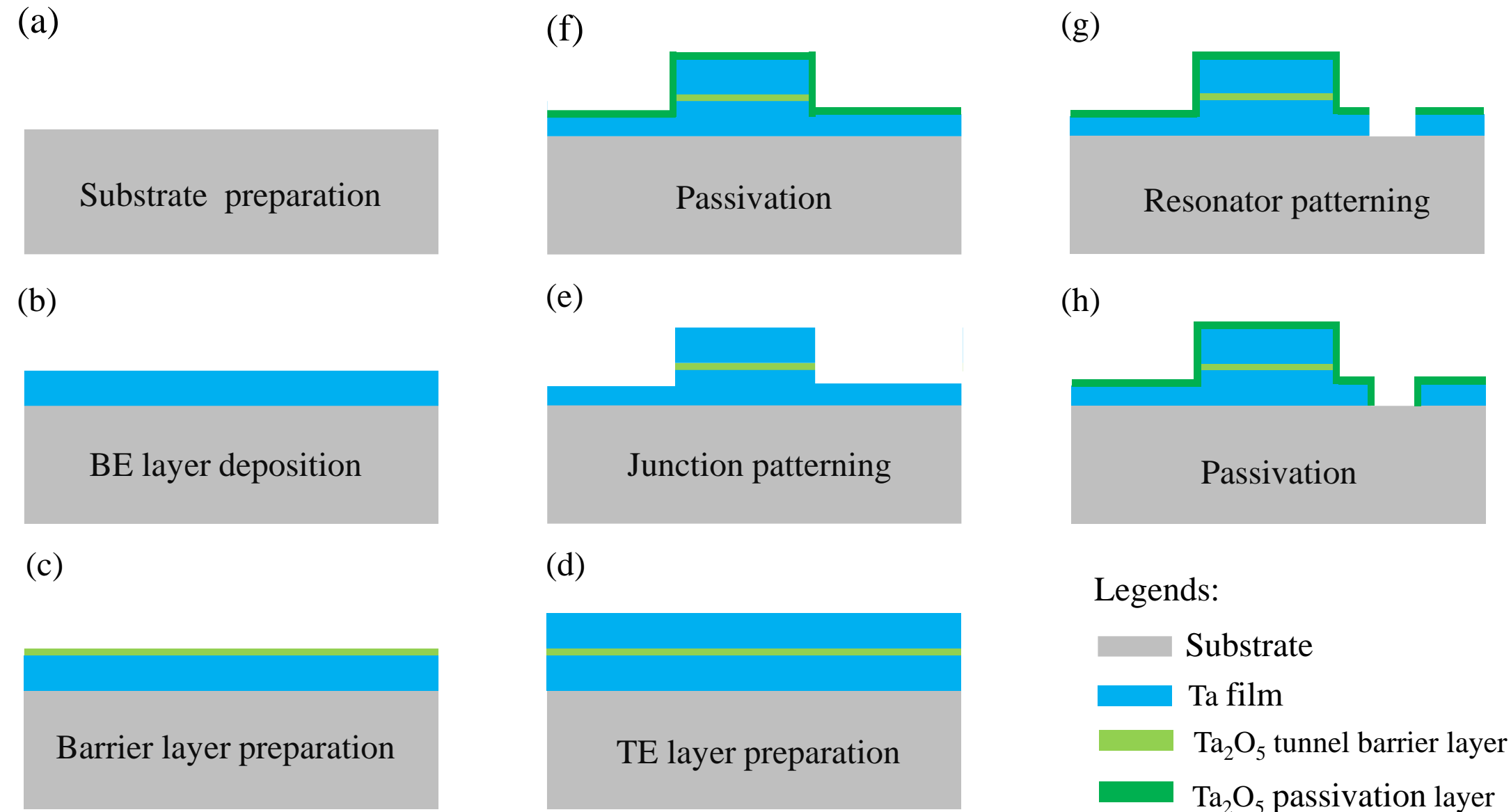

FIG.4. The process flow for the preparation of superconducting qubits based on $Ta_2O_5$ as dielectric and passivation layers. (a) Preparing the substrate; (b) Depositing α-Ta (110) film for the bottom electrode (BE) layer of Josephson junctions and other circuit structures; (c) Preparing the $Ta_2O_5$ tunnel barrier layer for the Josephson junctions; (d) Depositing α-Ta (110) film for the top electrode (TE) of Josephson junctions; (e) Patterning the junction area; (f) Surface passivation to obtain $Ta_2O_5$ passivation layer; (g) Patterning the resonators and other circuit structures, (h) Surface passivation to obtain $Ta_2O_5$ passivation layer.

## IV. CONCLUSIONS

In conclusion, we have investigated the surface oxides of α-Ta (110) films prepared in piranha. The oxide layer thickness on the α-Ta (110) surface is approximately 2.24 nm after exposing to air for four hours, containing various oxidization states of Ta. Treatment with the piranha solution primarily composed of pentavalent tantalum, with a thickness of approximately 2.61 nm. CPW resonators are fabricated using α-Ta (110) film with $Ta_2O_5$ surface oxide, exhibiting high $Q_i$ and robustness even after exposure to the atmosphere for months. Guided by these research results, we suggest the development of a full tantalum superconducting qubit, employing oxides with reduced dielectric loss and enhanced stability as passivation layers. This is consistent with the observations made in the fabrication of CPW resonators using α-Ta (110) film with $Ta_2O_5$ surface oxide.

## ACKNOWLEDGMENTS

K. L. X acknowledges support from the Youth Innovation Promotion Association of Chinese Academy of Sciences (2019319). J. G. F. acknowledges support from the Start-up foundation of Suzhou Institute of Nano-Tech and Nano-Bionics, CAS, Suzhou

(Y9AAD110).

## CONFLICT OF INTEREST

The authors have no conflicts to disclose.

## DATA AVAILABILITY

The data that support the findings of this study are available from the corresponding author upon reasonable request.